\documentclass[12pt]{article}
\title{New complex variables for equations \\ of ideal barotropic fluid}
\author{A.L. Sorokin \\
Institute of Thermophysics, 630090 Novosibirsk, Russia }
\begin{document}

\maketitle

\begin{abstract}
We propose new construction of dependent variables
for  equations of an ideal barotropic fluid. This construction
is based on a direct generalization of the known connection
between Schroedinger equation and a system of Euler-type equations.
The system for two complex-valued functions is derived that is
equivalent to Euler equations. Possible advantages of the proposed
formulation are discussed.
\end{abstract}

\section{Introduction}

When solving a partial problem of fluid dynamics
or exploring general properties of governing equations
one often use different choice of the dependent variables.
Introduction of a stream function is common practice
for two-dimensional problems. For a general case of a 3D
time-dependent flow one can use a vector potential, a pair of stream
functions (for incompressible case), Clebsch potentials and etc.
Clebsch potentials are mainly used with intention to exploit preferences
of Lagrange description of a fluid motion.
The new representation is based on the use
of multi-valued potentials and Euler approach.
The paper is composed as follows. In the second section we analyze
Madelung transformation that connects a generic Schroedinger equation
with a system of Euler-type equations. Some generalization
will be made for the case of potential flows of a barotropic fluid.
In the next section the generalization of Madelung transformation
for a general vector field will be derived, that leads to
the system of equations $(8)$ for two complex-valued functions with
arbitrary potentials.
In the fourth section we use this arbitrariness  and
propose the choice of potentials, that make the system equivalent
to Euler equations for an ideal barotropic fluid.
To substantiate this we will derive Euler equations from the system $(8)$.
In the last section we discuss possible preferences
of new choice of dependent variables and their relation
to vortices.

\section{Madelung transformation}

Since pioneer work by E.Madelung \cite{Mad26} physical literature
contains many examples of connection between Schroedinger equation
of quantum mechanics and fluid dynamics.
Typical exposition of this connection is the substitution
$\psi = \sqrt{\rho } \mathbf{e}^{i\frac {\varphi}{\beta}}$
into
\begin{equation}
i\frac{\partial \psi}{\partial t} = -\frac{\beta}{2}\Delta \psi + V\psi
\end{equation}
that leads to
\begin{equation}
\frac {\partial \rho } {\partial t}
+ \nabla \cdot \left( \rho \nabla \varphi \right) = 0
\end{equation}
\begin{equation}
\frac {\partial \varphi } {\partial t}
+ \frac {\left( \nabla \varphi \right)^2} {2}
= -V +
\frac {\beta} {2}
\frac {\Delta \sqrt \rho}{\sqrt \rho}
\end{equation}
This trick looks slightly mystical for novice.
Some historical notes and elucidation can be found in \cite{Sp80}.
More clear is back substitution. Following Madelung \cite{Mad57}, let's
linearize equation (2) using substitution
\begin{equation}
\rho = \psi \overline \psi,
\quad \varphi = -\frac{i\beta}{2}\ln \left( \frac{\psi}{\overline\psi} \right)
\end{equation}
were $\beta$ has dimension of kinematical viscosity.
After simple algebra one can obtain
$$
\overline \psi \left(
\frac{\partial \psi}{\partial t}
-\frac{i\beta}{2}\Delta \psi \right)
+ \psi \left(
\frac{\partial \overline \psi}{\partial t}
+\frac{i\beta}{2}\Delta \overline \psi \right)
=0
$$
Choice
$$
\frac{\partial \psi}{\partial t}
-\frac{i\beta}{2}\Delta \psi = iV\psi
$$
leads to Schroedinger equation.
Here $V$ is a real-valued function of a time, coordinates
and/or $\psi$.
We can conclude that this equation leads to conservation
of probability, but dynamics is completely defined by potential $V$.

Now from hydrodynamical viewpoint let's summarize restrictions
that were implicitly used in this derivation.
First, interpreting $\rho$ as density of some fluid with an arbitrary
equation of state, we see that fluid flow is supposed to be potential.
Second, we use dimensional constant $\beta$.

To describe an ideal fluid, we can to overcome
the second restriction using a non-dimensional form of equation (2)
$\left(\beta = 1\right)$
and the potential
$$
V=\Pi\left(\rho\right) +
\frac{1}{2}\frac {\Delta \sqrt \rho}{\sqrt \rho}
$$
This choice give Cauche-Lagrange equation for barotropic fluid
$$
\frac {\partial \varphi } {\partial t}
+ \frac {\left( \nabla \varphi \right)^2} {2}
= -\Pi
$$
but leads to
$$
i\frac{\partial \psi}{\partial t} =
-\frac{1}{4} \left(
\Delta\psi - \frac{\psi}{\overline\psi}\Delta\overline\psi
\right)
+\left[-\frac{1}{8}\left(\nabla\ln\left(\frac{\psi}{\overline\psi}\right)\right)^2
+\Pi\left(\psi\overline\psi\right)\right]\psi
$$
that differs from Schroedinger equation. This form of equation
of an ideal barotropic fluid seems to be unknown.

\section{Generalization of Madelung transformation}

We consider a direct generalization of the previous scheme for the case
of two complex-valued functions and introduce definitions
\begin{equation}
\rho =\rho_1 + \rho_2,\quad
\mathbf {J} =\rho\mathbf {V} = \rho_1 \nabla \varphi_1 + \rho_2 \nabla \varphi_2
\end{equation}
Obviously, permutation of indexes
should not have any physical consequence.
For velocity and vorticity we obtain
\begin{equation}
\mathbf {V} = \frac{\rho_1}{\rho} \nabla \varphi_1
+ \frac{\rho_2}{\rho} \nabla \varphi_2,\quad
\nabla \times \mathbf {V}=\frac{\rho_1\rho_2}{\rho^2}
\nabla\ln\left(\frac{\rho_1}{\rho_2}\right)
\times\nabla\left(\varphi_1-\varphi_2\right)
\end{equation}
The requirement of possibility to represent a vector field with
a non-zero total helicity
$$
H=\int{ \frac{\rho_1\rho_2}{\rho^2} \ln\left(\frac{\rho_1}{\rho_2}\right)
\left(\nabla\varphi_1\times\nabla\varphi_2\right)\cdot d\overline\sigma
}\ne 0
$$
implies a multi-valuedness of potentials \cite{Yok97}
(here integral should be taken over some closed surface).
That is admissible due to usage of the complex-valued variables.

Linearizing
\begin{equation}
\frac {\partial \rho } {\partial t}
+ \nabla \cdot \mathbf {J} = 0
\end{equation}
after some algebra we obtain
$$
\overline \psi_1
\left(\frac{\partial \psi_1}{\partial t}
-\frac{i}{2}\Delta \psi_1 \right)
+ \psi_1
\left(\frac{\partial \overline \psi_1}{\partial t}
+\frac{i}{2}\Delta \overline \psi_1 \right)
$$
$$
+\overline \psi_2 \left(
\frac{\partial \psi_2}{\partial t}
-\frac{i}{2}\Delta \psi_2 \right)
+ \psi_2 \left(
\frac{\partial \overline \psi_2}{\partial t}
+\frac{i}{2}\Delta \overline \psi_2 \right)
=0
$$
By inspection one can show that choice
$$
\frac{\partial \psi_k}{\partial t}
-\frac{i}{2}\Delta \psi_k
=U_k\psi_k
$$
with
$$
U_1=\frac{\rho_2}{2\rho}I-iV_1,\quad,U_2=-\frac{\rho_1}{2\rho}I-iV_2
$$
where $I,V_1,V_2$ are real-valued functions of time, coordinates
and/or $\psi_k$ solve this equation.
We obtain the following system of equations
\begin{equation}
i\frac{\partial \psi_1}{\partial t}
=-\frac{\Delta \psi_1}{2}
+\left(\frac{\rho_2}{2\rho}iI+V_1\right)\psi_1,
\quad
i\frac{\partial \psi_2}{\partial t}
=-\frac{\Delta \psi_2}{2}
+\left(-\frac{\rho_1}{2\rho}iI+V_2\right)\psi_2
\end{equation}
Substitutions $\psi_k=\sqrt\rho_k exp\left(i\varphi\right)$
give the  equivalent system
\begin{equation}
\frac {\partial \rho_k} {\partial t}
+ \nabla \cdot \left( \rho_k \nabla \varphi_k \right)
= \left(-1\right)^{k-1}\frac{\rho_1\rho_2}{\rho}I,
\quad
\frac {\partial \varphi_k } {\partial t}
+ \frac {\left( \nabla \varphi_k \right)^2} {2}
= -V_k +
\frac {1} {2}
\frac {\Delta \sqrt \rho_k}{\sqrt \rho_k}
\end{equation}

Equation (7) follows from the first two equations of this system.

\section{New form of Euler equations}

To apply the derived system to description of an ideal barotropic
flow we need a proper choice of the potentials $I,V_1,V_2$.
By inspection it was found that
\begin{equation}
V_1=\Pi\left(\rho\right)
- \frac{{\rho_2}^2}{2{\rho}^2}
\mathbf {w}^{2}
+
\frac {1} {2}
\frac {\Delta \sqrt \rho_1}{\sqrt \rho_1}
\end{equation}
\begin{equation}
V_2=\Pi\left(\rho\right) - \frac{{\rho_1}^2}{2{\rho}^2}
\mathbf {w}^{2}
+
\frac {1} {2}
\frac {\Delta \sqrt \rho_2}{\sqrt \rho_2}
\end{equation}
\begin{equation}
I=\nabla\cdot\mathbf {w}
+\frac{\mathbf {w}}{\rho }
\cdot\left(\rho _2\frac{\nabla\rho _1}{\rho _1}+
+\rho _1\frac{\nabla\rho _2}{\rho _2}\right)
\end{equation}
make system equivalent to Euler equations.
Here $\mathbf {w}=\nabla\left(\varphi _1-\varphi _2\right)$.
The invariance of systems (8),(9) with respect
to both Galilei group and indexes
permutation can be directly checked.

Substitution of (10-12) into (9) give
\begin{equation}
\frac {\partial \rho_1} {\partial t}
+ \nabla \cdot \left( \rho_1 \nabla \varphi_1 \right)
= \frac{\rho_1\rho_2}{\rho}I,
\quad
\frac {\partial \rho_2} {\partial t}
+ \nabla \cdot \left( \rho_2 \nabla \varphi_2 \right)
= -\frac{\rho_1\rho_2}{\rho}I,
\end{equation}
\begin{equation}
\frac {\partial \varphi_1 } {\partial t}
+ \frac {\left( \nabla \varphi_1 \right)^2} {2}
=-\Pi+\frac{{\rho_2}^2}{2{\rho}^2}\mathbf {w}^2
\end{equation}
\begin{equation}
\frac {\partial \varphi_2 } {\partial t}
+ \frac {\left( \nabla \varphi_2 \right)^2} {2}
=-\Pi+\frac{{\rho_1}^2}{2{\rho}^2}\mathbf {w}^2
\end{equation}

From equations (13) follows (7).

Now we start derivation of equation for flux $\mathbf {J}$.
First, multiplying (14),(15) by $\rho_k$ respectively,
summing and taking gradient of result, then adding
to obtained equation (13), multiplied by $\nabla\varphi_k$ respectively,
one can obtain
$$
\frac {\partial \mathbf {J} } {\partial t}
+ \nabla \left(  \frac { \mathbf {j_1}^2}{2\rho_1}
+\frac { \mathbf {j_2}^2}{2\rho_2} \right)
+\left[ \nabla\rho_1\frac{\partial \varphi_1} {\partial t}
+\nabla\rho_2\frac{\partial \varphi_2} {\partial t}\right]
$$
$$
+\left(\frac{\mathbf {j_1}\cdot\nabla\mathbf {j_1}}{\rho_1}
+\frac{\mathbf {j_2}\cdot\nabla\mathbf {j_2}}{\rho_2}
-\frac{\rho_1\rho_2}{\rho}I\mathbf {w}\right)
=-\nabla\left(\rho\Pi-\frac{\rho_1\rho_2}{\rho}\frac{\mathbf {w}^2}{2}\right)
$$
where $\mathbf {j_k}=\rho_k\nabla\varphi_k$.
Using identities
$$
\frac { \mathbf {J}^2}{2\rho}=
\frac { \mathbf {j_1}^2}{2\rho_1}+\frac { \mathbf {j_1}^2}{2\rho_1}
-\frac{\rho_1\rho_2}{\rho}\frac{\mathbf {w}^2}{2}
$$
$$
\frac{\mathbf {J}\nabla\cdot\mathbf {J}}{\rho}=
\frac{\mathbf {j_1}\nabla\cdot\mathbf {j_1}}{\rho_1}
+\frac{\mathbf {j_2}\nabla\cdot\mathbf {j_2}}{\rho_2}
-\frac{\rho_1\rho_2}{\rho}
\left(\frac{\nabla\cdot\mathbf {j_1}}{\rho_1}
-\frac{\nabla\cdot\mathbf {j_2}}{\rho_2}\right)\mathbf {w}
$$
after some algebra one can obtain
$$
\frac {\partial \mathbf {J} } {\partial t}
+ \nabla \left(  \frac { \mathbf {J}^2}{2\rho}\right)
-\frac { \mathbf {J}^2}{2\rho}\frac{\nabla\rho}{\rho}
+\frac{\mathbf {J}\nabla\cdot\mathbf {J}}{\rho}
$$
$$
+\left[
\nabla\rho_1\frac{\partial \varphi_1} {\partial t}
+\nabla\rho_2\frac{\partial \varphi_2} {\partial t}
+\frac { \mathbf {J}^2}{2\rho}\frac{\nabla\rho}{\rho}
+\Pi\nabla\rho
\right]
$$
$$
+\frac{\rho_1\rho_2}{\rho}
\left(
\frac{\nabla\cdot\mathbf {j_1}}{\rho_1}
-\frac{\nabla\cdot\mathbf {j_2}}{\rho_2}-I\right)\mathbf {w}
=-\rho\nabla\Pi
$$
Algebraic transformations of terms in square braces with account for
first identity and (14),(15) lead to equation
$$
\frac {\partial \mathbf {J} } {\partial t}
+ \nabla \left(  \frac { \mathbf {J}^2}{2\rho}\right)
-\frac { \mathbf {J}^2}{2\rho}\frac{\nabla\rho}{\rho}
+\frac{\mathbf {J}\nabla\cdot\mathbf {J}}{\rho}
$$
$$
+\frac{\rho_1\rho_2}{\rho}\left[
\nabla\ln\left(\frac{\rho_1}{\rho_2}\right)
\left(
\frac{\partial \varphi_1} {\partial t}
-\frac{\partial \varphi_2} {\partial t}
\right)
+
\left(
\frac{\nabla\cdot\mathbf {j_1}}{\rho_1}
-\frac{\nabla\cdot\mathbf {j_2}}{\rho_2}-I\right)\mathbf {w}
\right]
=-\rho\nabla\Pi
$$
Using definition of velocity and equations (14),(15)
after direct algebra one can show that terms in square braces
give Lamb vector
$$
\mathbf {V} \times \nabla \times \mathbf {V}
=\frac{\rho_1\rho_2}{\rho^2}
\left(
\left(\mathbf {V}\cdot\mathbf {w}\right)
\nabla\ln\left(\frac{\rho_1}{\rho_2}\right)
-\mathbf {V}\cdot\nabla\ln\left(\frac{\rho_1}{\rho_2}\right)\mathbf {w}
\right)
$$
We obtain the equation
\begin{equation}
\frac {\partial \mathbf {J} } {\partial t}
+ \nabla \left(  \frac { \mathbf {J} \cdot \mathbf {J}}{2\rho} \right)
- \frac { \mathbf {J} \cdot \mathbf {J}}{2\rho}
        \frac {\nabla \rho} {\rho}
- \mathbf {J} \times \nabla \times \mathbf {V}
+\mathbf {V} \nabla \cdot \mathbf {J}
=-\rho \nabla \Pi
\end{equation}
To make last step in derivation one should use continuity equation
to obtain from (16) Euler equation in Gromeka-Lamb form
\begin{equation}
\frac {\partial \rho } {\partial t}
+ \nabla \cdot \mathbf {J} = 0,
\quad
\frac {\partial \mathbf {V} } {\partial t}
+ \nabla \left( \frac { \mathbf {V} \cdot \mathbf {V}}{2} \right)
- \mathbf {V} \times \nabla \times \mathbf {V}
=-\nabla \Pi
\end{equation}
The result is as follows: System (8) is equivalent
to system of Euler equation (17).

\section{Discussion}

First of all, the attractive feature of (8) is the homogeneity
both dependent variables and equations
in contrast to the non-homogeneity  of velocity/density  and form
of equations in (18).
This property can be used both numerically and analytically.
Homogeneity and elimination of the convective derivative
can substantially simplify numerical algorithm.
As far as multivaluedness is concerned, the possibility of use
multi-valued potentials was clearly demonstrated in \cite{Bra95}.
In analytical way the aforementioned property can simplify
proof of existence and uniqueness theorems.
Also application of geometrical methods to partial differential equations (8)
is looking quite natural.

This formulation of Euler equation can have another interesting property.
Zeroes of solution of nonlinear Schoedinger equation
correspond to a vortex axes (topological defects) \cite{Bra95}.
At a moment the condition $\psi=0$ defines two surfaces,
and their intersection defines a space curve (possibly,
disconnected).
Note similarity with definition of a vortex as zero
of an analitical complex-valued function in two-dimensional
hydrodynamic of ideal incompressible fluid.
If the system (8) inherits this property from its prototype (1)
the known problem of a vortex definition \cite{Hus95}
can be solved in general case.

\section{Acknowledgments}

Author express his gratitude to Prof. S.K.Nemirovsky and Dr. G.A.Kuz'min.

\bibliographystyle{unsrt}
\bibliography{bib}

\end{document}